\documentclass[aps,pra,preprint,groupedaddress]{revtex4}
\usepackage{graphicx}
\usepackage{amsmath}

\begin{document}

\title{Quantum Error Models and Error Mitigation \\
        for Long-Distance Teleportation Architectures}

\author{Jeffrey H. Shapiro}
\author{Joe Aung}
\author{Brent J. Yen}
\affiliation{Research Laboratory of Electronics \\
              Massachusetts Institute of Technology, Cambridge, MA 02139}

\date{\today}

\begin{abstract}

A quantum communication architecture is being developed for
long-distance, high-fidelity qubit teleportation.  It uses
an ultrabright narrowband source of polarization-entangled photons, plus
trapped-atom quantum memories, and it is compatible with long-distance
transmission over standard telecommunication fiber.  This paper reports
error models for the preceding teleportation architecture, and for an
extension thereto which enables long-distance transmission and
storage of Greenberger-Horne-Zeilinger states.  The use of
quantum error correction or entanglement purification to improve the
performance of these quantum communication architectures is also
discussed.
\end{abstract}

\maketitle

\section{Introduction}

     A team of researchers from the Massachusetts Institute of
Technology (MIT) and Northwestern University (NU) has proposed a
quantum communication architecture \cite{Arch} that permits
long-distance high-fidelity teleportation using the Bennett et al.
singlet-state protocol \cite{teleportation}.  This architecture
uses a novel ultrabright source of polarization-entangled photon
pairs \cite{ultra} and trapped-atom quantum memories
\cite{atommemory} in which all four Bell states can be measured.
By means of quantum-state frequency conversion and time-division
multiplexed polarization restoration, it is able to employ
standard telecommunication fiber for long-distance transmission of
the polarization-entangled photons. An extension of the MIT/NU
architecture enables long-distance transmission and storage of the
Greenberger-Horne-Zeilinger (GHZ) states that are needed for
quantum secret sharing protocols. In this paper, we report
quantum-communication error models for the MIT/NU architectures,
and we describe the use of quantum error correction or
entanglement purification to improve the robustness
of these quantum transmission systems.

\section{\label{MITNU} MIT/NU Communication Architecture}

The notion that singlet states could be used to teleport a qubit is due
to  Bennett et al.\@ \cite{teleportation}.  The transmitter and the
receiver stations share the entangled qubits of a singlet state,
$|\psi^-\rangle_{TR} = (|0\rangle_T|1\rangle_R -
|1\rangle_T|0\rangle_R)/\sqrt{2}$, and the transmitter then accepts a
message qubit, $|\psi\rangle_M = \alpha|0\rangle_M +
\beta|1\rangle_M$, leaving the message mode, the transmitter, and the
receiver in the joint state $|\psi\rangle_M|\psi^-\rangle_{TR}$.  Making
the Bell-state measurements,
$\{|\psi^{\pm}\rangle_{MT} = (|1\rangle_M|0\rangle_T \pm
|0\rangle_M|1\rangle_T)/\sqrt{2},
|\phi^{\pm}\rangle_{MT} = (|1\rangle_M|1\rangle_T \pm
|0\rangle_M|0\rangle_T)/\sqrt{2}\}$, on the joint message/transmitter
system then yields the two bits of classical information that the
receiver needs to transform its portion of the original singlet into
a reproduction of the message qubit.

\begin{figure}
\includegraphics[width=2.5in]{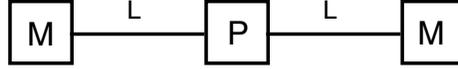}
\caption{\label{arch} Schematic of long-distance quantum
communication system: $P$ = ultrabright narrowband source of
polarization-entangled photon pairs; $L$ = $L$\,km of standard
telecommunications fiber; $M$ = trapped-atom quantum memory.}
\end{figure}

An initial experimental
demonstration of teleportation using singlet states was performed by
Bouwmeester et al.\@
\cite{bouw1},\cite{bouw2}, but only one of the Bell states was measured,
the demonstration was a table-top experiment, and it did not include a
quantum memory.  The MIT/NU proposal for a singlet-based quantum
communication system, which is shown in Fig.~\ref{arch}, remedies all of
these limitations.  It uses an ultrabright source of
polarization-entangled photon pairs, formed by combining the outputs from
two coherently-pumped, type-II phase-matched optical parametric
amplifiers on a polarizing beam splitter.  It transmits one photon from
each pair down standard telecommunication fibers to a pair of
trapped $^{87}$Rb-atom quantum memories for storage and processing of
this entanglement.  One of these memories serves as the transmitter
station and the other as the receiver station for qubit teleportation.  We
will devote the rest of this section to describing these basic components
and their operation within the MIT/NU quantum communication architecture.

\subsection{Ultrabright Source of Polarization-Entangled Photons}

\begin{figure}
\includegraphics[width=4.5in]{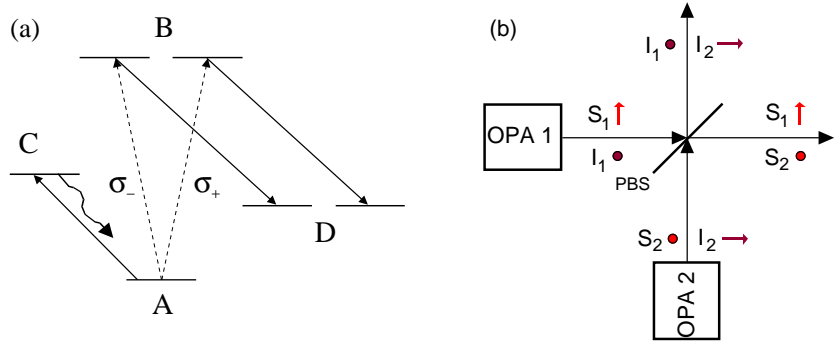}
\caption{\label{sourceandmemory} Essential components of the
singlet-state quantum communication system from Fig.~1. (a) Energy
level diagram of the trapped rubidium atom quantum memory. The
$A$-to-$B$ transition occurs when a photon is absorbed. The
$B$-to-$D$ transition is coherently driven to enable storage in
the long-lived $D$ levels. The $A$-to-$C$ cycling transition is
used for nondestructive verification of a loading event. (b)
Ultrabright narrowband source of polarization-entangled photon
pairs. The polarizations $\hat{x}$ and $\hat{y}$ are denoted by
arrows and bullets, respectively; PBS=polarizing beam splitter.}
\end{figure}

Polarization-entangled photons are transmitted from the source
over $L$\,km of standard optical fiber to be loaded in trapped-atom
quantum memories.  The quantum memories in the system require a
source of entangled photons at 795\,nm to match the cavity
linewidth of the trapped-atom cavities. The standard source for
generating polarization-entangled photons is the parametric
downconverter.  It is so broadband ($\sim$$10^{13}$\,Hz), however, that
its  pair-generation rate in the narrow bandwidth needed for coupling
into the rubidium atom is extremely low:    $\sim$15\,pairs/sec in a
30\,MHz bandwidth.  The $P$ block in Fig.~\ref{arch} represents an
ultrabright narrowband source \cite{ultra}, which is capable of producing
$1.5 \times 10^{6}$\,pairs/sec in a 30\,MHz bandwidth by combining the
signal and idler output beams from two doubly resonant type-II phase
matched OPAs, as sketched in Fig.~\ref{sourceandmemory}(b).

\begin{figure}
\includegraphics[width=3.5in]{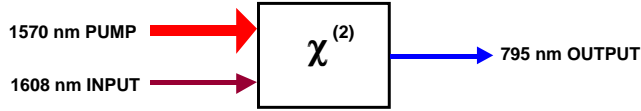}
\caption{\label{upconvert} Schematic diagram of quantum-state frequency
conversion: a strong pump beam at 1570\,nm converts a qubit photon
received at 1608\,nm (in the low-loss fiber transmission window) to
a qubit photon at the 795\,nm wavelength of the $^{87}$Rb quantum
memory.}
\end{figure}

Quasi-phase-matched nonlinear materials make it possible to realize a
wavelength, for our polarization-entanglement source, that is appropriate
for a specific application.  In particular, by using periodically-poled
potassium titanyl phosphate (PPKTP), a quasi-phase-matched type-II
nonlinear material, we can produce
$\sim$$10^6$\,pairs/sec at the 795\,nm wavelength of the rubidium
memory for direct memory-loading (i.e., local-storage)
applications.  For long-distance transmission to remotely located
memories, we can use a different PPKTP crystal and pump wavelength to
generate $10^6$\,pairs/sec in the 1.55\,$\mu$m wavelength low-loss
fiber transmission window. After fiber propagation we shift the
entanglement to the 795\,nm wavelength needed for the rubidium-atom
memory via quantum-state frequency conversion \cite{kumar},\cite{huang},
shown in Fig.~\ref{upconvert}.

\subsection{Quantum-State Transmission over Fiber}

\begin{figure}
\includegraphics[width=4in]{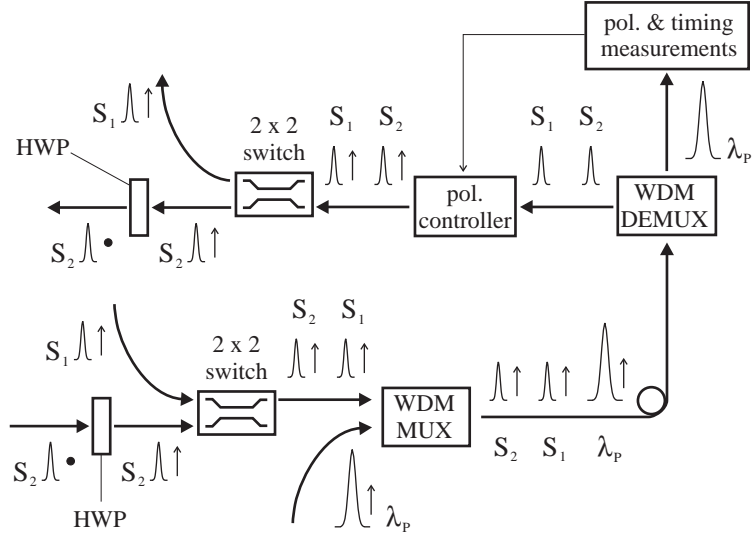}
\caption{\label{tdm} Transmission of time-division multiplexed
signal beams from OPAs~1 and 2 through an optical fiber.
$\lambda_p$ = pilot pulse, WDM MUX = wavelength-division
multiplexer, WDM DEMUX = wavelength-division demultiplexer, HWP =
half-wave plate.}
\end{figure}

Successful singlet transmission requires that polarization not be
degraded by the propagation process.  Yet, propagation through standard
telecommunication fiber produces random, slowly-varying ($\sim$msec
time scale) polarization variations, so a means for polarization
restoration is required.  The approach taken for polarization restoration
in the MIT/NU architecture, shown schematically
in Fig.~\ref{tdm}, relies on time-division multiplexing (TDM). Time slices
from the signal beams from the two OPAs are sent down one fiber in the
same linear polarization but in nonoverlapping time slots, accompanied by
a strong out-of-band pulse.  By tracking and restoring the linear
polarization of the strong pulse, we can restore the linear polarization
of the signal-beam time slices at the far end of the fiber. After this
linear-polarization restoration, we then reassemble a time-epoch of the
full vector signal beam by delaying the first time slot and combining it
on a polarizing beam splitter with the second time slot after the latter
has had its linear polarization rotated by $90^{\circ}$.  A similar
procedure is performed to reassemble idler time-slices after they have
propagated down the other fiber. This approach, which is inspired
by the Bergman et al.\@ two-pulse fiber-squeezing experiment
\cite{bergman}, common-modes out the vast majority of the phase
fluctuations and the polarization birefringence incurred in the
fiber, permitting standard telecommunication fiber to be used in
lieu of the lossier and much more expensive
polarization-maintaining fiber.

\subsection{Trapped-Atom Quantum Memory}

Each $M$ block in Fig.~\ref{arch} is a quantum memory in which a
single ultra-cold $^{87}\text{Rb}$ atom ($\sim$6\,MHz linewidth) is
confined by a far-off-resonance laser trap in an ultra-high-vacuum
chamber with cryogenic walls within a high-finesse ($\sim$15\,MHz
linewidth) single-ended optical cavity. This memory can absorb a
795\,nm photon, in an arbitrary polarization state, transferring
the qubit from the photon to the degenerate $B$ levels of
Fig.~\ref{sourceandmemory}(a) and thence to long-lived storage
levels, by coherently driving the $B$-to-$D$ transitions.  (We are
using abstract symbols here for the hyperfine levels of rubidium;
see \cite{atommemory} for the actual atomic levels involved as
well as a complete description of the memory and its operation.)
With a liquid helium cryostat, so that the background pressure is
less than $10^{-14}$\,Torr, the expected lifetime of the trapped
rubidium atom will be more than an hour. Fluctuations in the
residual magnetic field, however, will probably limit the atom's
decoherence time to a few minutes.

By using optically off-resonant Raman (OOR) transitions, the Bell
states of two atoms in a single vacuum-chamber trap can be
converted to superposition states of one of the atoms.  All four
Bell measurements can then be made, sequentially, by detecting the
presence (or absence) of fluorescence as an appropriate sequence
of OOR laser pulses is applied to the latter atom
\cite{atommemory}. The Bell-measurement results in one memory can
be sent to a distant memory, where at most two additional OOR
pulses are needed to complete the Bennett et al.\@ state
transformation.  The qubit stored in a trapped rubidium atom can
be converted back into a photon by reversing the Raman excitation
process that occurs during memory loading.

\subsection{\label{memoryloading}Memory Loading Protocol}

The MIT/NU quantum communication system is clocked, with the following
memory loading protocol being run every cycle.  Time slots of signal
and idler photons are transmitted over optical fibers in the 1.55\,$\mu$m
low-loss window, upconverted to 795\,nm, and gated into their respective
quantum memories. During a short cavity-loading interval of a few
cold-cavity lifetimes, say 400\,ns, the atoms are optically detuned or
physically displaced to prevent
$A$-to-$B$ absorptions of 795\,nm photons.  After this loading interval,
the atoms are tuned or moved into absorbing positions and the
$B$-to-$D$ transition is coherently pumped for 100\,ns.  To
test whether each memory has loaded a photon in its
$D$ storage levels---without destroying the coherences stored therein---we
repeatedly drive each atom's
$A$-to-$C$ transition and monitor both cavities for
fluorescence from these cycling transitions.  If it is determined that
either atom has failed to absorb a 795\,nm photon, i.e., if either atom
has failed to be excited into a superposition of its long-lived
$D$ levels, then both atoms are returned to their $A$ states and
the loading protocol is repeated.  We expect that this memory-loading
protocol could be run at rates as high as $R$ = 500\,kHz, so that we
can attempt to load an entangled photon pair once every 2\,$\mu$s.

\section{Teleportation-System Error Model}

The performance of the MIT/NU architecture was studied in
Ref.~\cite{Arch}, using a simple model that assumes all error
events lead to storage of independent random polarizations.  In
this section, we present a more accurate error model, and we discuss the
use of quantum error correction or entanglement purification
to obtain improved system performance.

Let $\hat{a}_k$, $k = S_x, S_y, I_x, I_y$, be the annihilation
operators of the optical field modes within the quantum memory cavities
at the end of a cold-cavity load.  It was shown in
\cite{Arch} that the joint density operator for these modes takes the
factored form, $\hat{\rho}_{\mathbf{SI}} =
\hat{\rho}_{SxIy}\otimes\hat{\rho}_{SyIx}$, where the two-mode
density operators on the right-hand side are Gaussian mixed states
given by the anti-normally ordered characteristic functions,
\begin{align}
\label{twomode1}
    \chi_A^{\rho_{S_xI_y}}(\zeta_S,\zeta_I)
        &\equiv  \text{tr}\left(
                \hat{\rho}_{S_xI_y}
                e^{-\zeta^{*}_S\hat{a}_{S_x}
                   -\zeta^{*}_I\hat{a}_{I_y}}
                e^{\zeta_S\hat{a}^{\dagger}_{S_x}
                   +\zeta_I\hat{a}^{\dagger}_{I_y}} \right)  \\
        &=  \exp\left[-(1+\bar{n})(|\zeta_S|^2 + |\zeta_I|^2)
                  + 2\tilde{n}\text{Re}(\zeta_S\zeta_I)
             \right],
\end{align}
and
\begin{align}
\label{twomode2}
    \chi_A^{\rho_{S_yI_x}}(\zeta_S,\zeta_I)
        &\equiv  \text{tr}\left(
                \hat{\rho}_{S_yI_x}
                e^{-\zeta^{*}_S\hat{a}_{S_y}
                   -\zeta^{*}_I\hat{a}_{I_x}}
                e^{\zeta_S\hat{a}^{\dagger}_{S_y}
                   +\zeta_I\hat{a}^{\dagger}_{I_x}} \right)  \\
        &=  \exp\left[-(1+\bar{n})(|\zeta_S|^2 + |\zeta_I|^2)
                  - 2\tilde{n}\text{Re}(\zeta_S\zeta_I)
             \right],
\end{align}
with $\bar{n} = I_{-} - I_{+}, \tilde{n} = I_{-} + I_{+}$ and
$I_{\pm} \equiv \eta_L\gamma\gamma_c G/[\Gamma\Gamma_c(1\pm G)(1\pm G +
\Gamma_c/\Gamma)]$. Here: $G^2$ is the normalized OPA pump power ($G^2=1$
at oscillation threshold);
$\eta_L$ is the propagation loss; $\gamma$ and $\gamma_c$ are the
output-coupling rates of the OPA and the memory cavities; and $\Gamma$
and $\Gamma_c$ are the linewidths of the OPA and memory cavities.
The joint density operator for the $S_x$ and $I_y$ modes can be found from
the inverse relation for
\eqref{twomode1}, viz.,
\begin{equation}
\label{inverse}
  \hat{\rho}_{S_xI_y} =
                       \int\frac{d^2\zeta_S}{\pi}
                       \int\frac{d^2\zeta_I}{\pi}
                       \chi_A^{\rho_{S_xI_y}}(\zeta_S,\zeta_I)
                       e^{-\zeta_S\hat{a}^{\dagger}_{S_x}
                          -\zeta_I\hat{a}^{\dagger}_{I_y}}
                       e^{\zeta^{*}_S\hat{a}_{S_x}
                         +\zeta^{*}_I\hat{a}_{I_y}};
\end{equation}
a similar inverse relation exists for \eqref{twomode2}. These
expressions will be used to derive an error model for our quantum
communication system.

\subsection{\label{singlephoton} Single-Photon Event Model}

In Section~\ref{memoryloading} we described a procedure for
nondestructively detecting whether a quantum memory has absorbed a
photon.  This procedure allows us to isolate erasure events, i.e.,
loading intervals in which one or both atoms fail to absorb photons.
Erasures reduce the throughput, but they do not reduce the teleportation
fidelity achieved by the MIT/NU architecture.  Because OPA sources can
produce more than one pair in a loading interval, we are not guaranteed
that the desired singlet state has been loaded into the two quantum
memories when no cycling fluorescence is detected from either memory
during a loading interval.  Indeed, it is possible that two pairs were
emitted by the source, with one photon from the first pair being lost
en route down one fiber, and one photon from the second pair being lost
while in propagation down the other fiber.  In this situation, the two
atoms may not be placed in a singlet state,  hence an error may be
incurred in subsequent a teleportation procedure that uses the coherences
that get stored in these atoms.   The cold-cavity loading analysis
presented in
\cite{Arch} calculated the erasure probability, i.e., the probability
that one or both of the memory cavities failed to absorb a photon after a
trial of the loading protocol, and the success probability, i.e., the
probability that the two memory cavities share a singlet state after a
trial of the loading protocol.  All other possibilities were considered
to be errors, and assumed to leave  the memories in independent states of
random polarization.  Inasmuch as it is possible---in the cold-cavity
loading analysis---for one or both of the memory cavities to absorb \em
more\/\rm\ than one photon, the error probability calculated in
\cite{Arch} includes multiphoton errors, i.e., error events in which both
memory cavities absorb photons with at least one of them absorbing two or
more photons.  We now show how these multiphoton events can be eliminated
from the error event.

\begin{figure}
\includegraphics[width=3in]{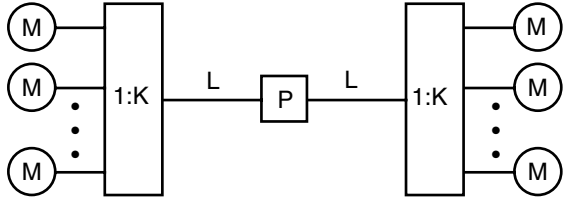}
\caption{\label{1_M_arch} Conversion of multiphoton events into
erasures. Beam splitter arrays are used to achieve a 1:$K$ fanout at
the far end of each optical fiber.  The $K$ outputs from each array are
connected to single-rubidium atom quantum memories.   An erasure is
declared unless exactly
one memory at each end has loaded a photon.}
\end{figure}

Consider the configuration shown in Fig.~\ref{1_M_arch}.  Here, multiple
memories convert multiphoton events into erasures.   We use beam-splitter
arrays to achieve 1:$K$ equal-splitting fanouts at the far end of each
optical fiber.  The $K$ outputs from each array are connected to $K$
single-rubidium atom quantum memories. During each loading  interval,
\em all\/\rm\ of the quantum memories are monitored for
$A$-to-$C$ fluorescence.  An erasure is declared---and the
memory loading protocol is repeated---unless exactly one memory at each
end has absorbed a photon.  In the limit of large
$K$, this scheme converts all multiphoton events into erasures.  The only
loading events that remain are those in which a single photon entered
exactly one of the memories at each end of the quantum communication
system.  These loaded memories are then the pair that is used to perform
teleportation.  An error now occurs when the two loaded memories are not
in the singlet state.  Because only one photon was absorbed at each
memory, all possible loading events can be represented in the Bell
basis,
$\{|\psi^{\pm}\rangle_{TR} = (|0\rangle_T|1\rangle_R \pm
|1\rangle_T|0\rangle_R)/\sqrt{2},
|\phi^{\pm}\rangle_{TR} = (|1\rangle_T|1\rangle_R \pm
|0\rangle_T|0\rangle_R)/\sqrt{2}\}$, where
\begin{align}
|0\rangle_T &\equiv|1\rangle_{S_x}|0\rangle_{S_y}\quad\mbox{and}\quad
|1\rangle_T \equiv |0\rangle_{S_x}|1\rangle_{S_y},\\
|0\rangle_R &\equiv |1\rangle_{I_x}|0\rangle_{I_y}\quad\mbox{and}\quad
|1\rangle_R \equiv |0\rangle_{I_x}|1\rangle_{I_y},
\end{align}
define the logical qubits at the transmitter and the receiver in terms of
their respective cavity-field-mode states.  We shall characterize
the joint density operator for single-photon loading events by finding
closed-form expressions for its Bell-basis matrix elements.

We first compute the diagonal entries of the density matrix. The
probability of loading the singlet state
$|\psi^{-}\rangle_{TR}$
is,
\begin{align}
    \Pr(|\psi^{-}\rangle_{TR})
             &= \text{$_{TR}\langle \psi^{-}|$}
                \hat{\rho}_{\bf{SI}}
                \text{$|\psi^{-}\rangle_{TR}$} \\
             &= \frac{1}{2} \left(
                _{S_xI_y}\langle 00|
                \hat{\rho}_{S_xI_y}
                |00\rangle_{S_xI_y} \,
                {}_{S_yI_x}\langle 11|
                \hat{\rho}_{S_yI_x}
                |11\rangle_{S_yI_x} \right. \notag \\
            & \quad \; + \;
                _{S_xI_y}\langle 11|
                \hat{\rho}_{S_xI_y}
                |11\rangle_{S_xI_y} \,
                {}_{S_yI_x}\langle 00|
                \hat{\rho}_{S_yI_x}
                |00\rangle_{S_yI_x} \notag \\
            & \quad\; - \;
                _{S_xI_y}\langle 00|
                \hat{\rho}_{S_xI_y}
                |11\rangle_{S_xI_y} \,
                {}_{S_yI_x}\langle 11|
                \hat{\rho}_{S_yI_x}
                |00\rangle_{S_yI_x}  \notag \\
           & \quad\; - \; \left.
                _{S_xI_y}\langle 11|
                \hat{\rho}_{S_xI_y}
                |00\rangle_{S_xI_y} \,
                {}_{S_yI_x}\langle 00|
                \hat{\rho}_{S_yI_x}
                |11\rangle_{S_yI_x} \right) \\
           &= p_{00}p_{11} + p_{c},
\end{align}
where
\begin{align}
p_{00} &=
{}_{S_xI_y}\langle 00|\hat{\rho}_{S_xI_y}|00\rangle_{S_xI_y}
= {}_{S_yI_x}\langle 00|\hat{\rho}_{S_yI_x}|00\rangle_{S_yI_x},
\label{p00}\\
p_{11} &=
{}_{S_xI_y}\langle 11|\hat{\rho}_{S_xI_y} |11\rangle_{S_xI_y}
= {}_{S_yI_x}\langle 11|\hat{\rho}_{S_yI_x} |11\rangle_{S_yI_x},
\label{p11}\\
p_{c} &=
|{}_{S_xI_y}\langle 00| \hat{\rho}_{S_xI_y}|11\rangle_{S_xI_y}|^2
= |{}_{S_yI_x}\langle 00| \hat{\rho}_{S_yI_x}|11\rangle_{S_yI_x}|^2,
\label{pc}
\end{align}
with the second equalities in Eqs.~\eqref{p00}--\eqref{pc} following
from comparing the anti-normally ordered characteristic function
for the $\{S_x,S_y\}$ modes to that for the $\{I_x,I_y\}$ modes.
A similar calculation for the probabilities of
the triplet states shows that,
\begin{align}
  \Pr(|\psi^{+}\rangle_{TR}) &= p_{00}p_{11} - p_{c}, \\
  \Pr(|\phi^{-}\rangle_{TR}) &= p_{10}^2, \\
  \Pr(|\phi^{+}\rangle_{TR}) &= p_{10}^2,
\end{align}
where
\begin{align}
p_{10} &=
|{}_{S_xI_y}\langle 10|\hat{\rho}_{S_xI_y} |10\rangle_{S_xI_y}|^2 =
|{}_{S_xI_y}\langle 01|\hat{\rho}_{S_xI_y} |01\rangle_{S_xI_y}|^2
\notag \\
&=
|{}_{S_yI_x}\langle 10|\hat{\rho}_{S_yI_x} |10\rangle_{S_yI_x}|^2 =
|{}_{S_yI_x}\langle 01|\hat{\rho}_{S_yI_x} |01\rangle_{S_yI_x}|^2.
\end{align}

To evaluate $\{p_{00},p_{10},p_{11},p_{c}\}$,
we parallel the technique used in
\cite{Arch} for computing similar quantities.  We observe that
\begin{equation}
    \chi^{\rho_{S_xI_y}}_A(\boldsymbol{\zeta})
    = \frac{\pi^2
    p_{S_xI_y}(\boldsymbol{\zeta})}{(1+\bar{n})^2 - \tilde{n}^2},
\end{equation}
where $p_{S_xI_y}(\boldsymbol{\zeta})$ is the classical
probability density for a zero-mean, complex-valued Gaussian
random vector $\boldsymbol{\zeta}^T = [\zeta_S \;\; \zeta_I] $
with second-moment matrices
\begin{equation}
   \langle \boldsymbol{\zeta}
           \boldsymbol{\zeta}^{\dagger} \rangle_{p_{S_xI_y}}
   =  \frac{1}{(1+\bar{n})^2 - \tilde{n}^2}
     \begin{bmatrix} 1+\bar{n} & 0 \\
                    0 & 1+\bar{n}
     \end{bmatrix},
\end{equation}
and
\begin{equation}
   \langle \boldsymbol{\zeta}
           \boldsymbol{\zeta}^T \rangle_{p_{S_xI_y}}
   = \frac{1}{(1+\bar{n})^2 - \tilde{n}^2}
     \begin{bmatrix} 0 & \tilde{n} \\
                     \tilde{n} & 0
     \end{bmatrix}.
\end{equation}
Using the inverse relation \eqref{inverse} for the density
operator, we find:
\begin{equation}
   p_{00} =  {}_{S_xI_y}\langle 00|
                \hat{\rho}_{S_xI_y}
                |00\rangle_{S_xI_y}
          = \iint \frac{d^2\zeta_S d^2\zeta_I}{(1+\bar{n})^2 - \tilde{n}^2}
             p_{S_xI_y}(\boldsymbol{\zeta})
          = \frac{1}{(1+\bar{n})^2 - \tilde{n}^2},
\end{equation}
\begin{align}
   p_{10} =  {}_{S_xI_y}\langle 10|
                \hat{\rho}_{S_xI_y}
                |10\rangle_{S_xI_y}
          &= \iint \frac{d^2\zeta_S d^2\zeta_I}{(1+\bar{n})^2 - \tilde{n}^2}
             (1 - |\zeta_S|^2)
             p_{S_xI_y}(\boldsymbol{\zeta}) \\
          &= \frac{1 - \langle |\zeta_S|^2 \rangle_{p_{S_xI_y}}}{(1+\bar{n})^2 -
             \tilde{n}^2}  \\
          &= \frac{\bar{n}(1 + \bar{n}) - \tilde{n}^2}
                  {[(1+\bar{n})^2 - \tilde{n}^2]^2},
\end{align}
\begin{align}
   p_{11} =  {}_{S_xI_y}\langle 11|
                \hat{\rho}_{S_xI_y}
                |11\rangle_{S_xI_y}
          &= \iint \frac{d^2\zeta_S d^2\zeta_I}{(1+\bar{n})^2 - \tilde{n}^2}
             (1 - |\zeta_S|^2)(1 - |\zeta_I|^2)
             p_{S_xI_y}(\boldsymbol{\zeta}) \\
          &= \frac{\langle(1 - |\zeta_S|^2)(1 - |\zeta_I|^2) 
\rangle_{p_{S_xI_y}}}{(1+\bar{n})^2 -
             \tilde{n}^2}  \\
          &= \frac{(\bar{n}(1 + \bar{n}) - \tilde{n}^2)^2 + \tilde{n}^2}
          {[(1+\bar{n})^2 - \tilde{n}^2]^3} \label{momentfactoring},
\end{align}
and
\begin{align}
   p_{c} =
               |_{S_xI_y}\langle 00|
                \hat{\rho}_{S_xI_y}
                |11\rangle_{S_xI_y}|^2
          &= \left|\iint \frac{d^2\zeta_S d^2\zeta_I}{(1+\bar{n})^2 - 
\tilde{n}^2}
             \zeta_S \zeta_I
             p_{S_xI_y}(\boldsymbol{\zeta}) \right|^2 \\
          &= \left| \frac{\langle \zeta_S \zeta_I 
\rangle_{p_{S_xI_y}}}{(1+\bar{n})^2 -
             \tilde{n}^2} \right|^2  \\
          &= \frac{\tilde{n}^2}
          {[(1+\bar{n})^2 - \tilde{n}^2]^4}.
\end{align}
Equation~\eqref{momentfactoring} follows from the moment factoring
theorem for complex-valued Gaussian random variables.  The
off-diagonal entries of the density matrix are computed in the
same way, and it turns out that they are all zero. (See
Ref.~\cite{jaung} for details of these calculations.) Note that we
can verify from the expressions above that $p_{00}p_{11} -
p_{c} = p_{10}^2$, i.e., the triplet state
probabilities are all equal.

Normalizing the preceding matrix elements to set
$\mbox{tr}(\hat{\rho}_{TR}) = 1$, we obtain the conditional
density operator for single-photon loading events given that an erasure
did \em not\/\rm\ occur.  This conditional density operator
for the Fig.~\ref{1_M_arch} architecture is diagonal in the Bell basis,
and given by,
\begin{equation}
\label{werner}
\hat{\rho}_{TR} =
      \text{diag}\!\begin{bmatrix}
        P_s & (1-P_s)/3 & (1-P_s)/3 & (1-P_s)/3
      \end{bmatrix},
\end{equation}
where $P_s \equiv
[(\bar{n}(1+\bar{n}) - \tilde{n})^2 +
2\tilde{n}^2]/[4(\bar{n}(1+\bar{n}) - \tilde{n})^2 +
2\tilde{n}^2]$, is the conditional probability of loading a singlet given
there has not been an erasure, i.e., it is the conditional success
probability.  Equation~\eqref{werner} is a
$|\psi^-\rangle_{TR}$ Werner state, so that teleporting the message qubit
$\hat{\rho}_M$ from transmitter $T$ to receiver $R$ using the entangled
mixed state
$\hat{\rho}_{TR}$ is equivalent to transmitting $\hat{\rho}_M$ over a
depolarizing quantum channel of fidelity $P_s$.
The average teleportation fidelity $F$ realized with the Bennett et al.\@
protocol, assuming the input qubit is a random pure state chosen from a
uniform distribution over the Bloch sphere, is easily shown from
$\hat{\rho}_{TR}$ to be,
\begin{equation}
     F = (2P_s+1)/3.
\end{equation}
\begin{figure}
\includegraphics[width=2.75in]{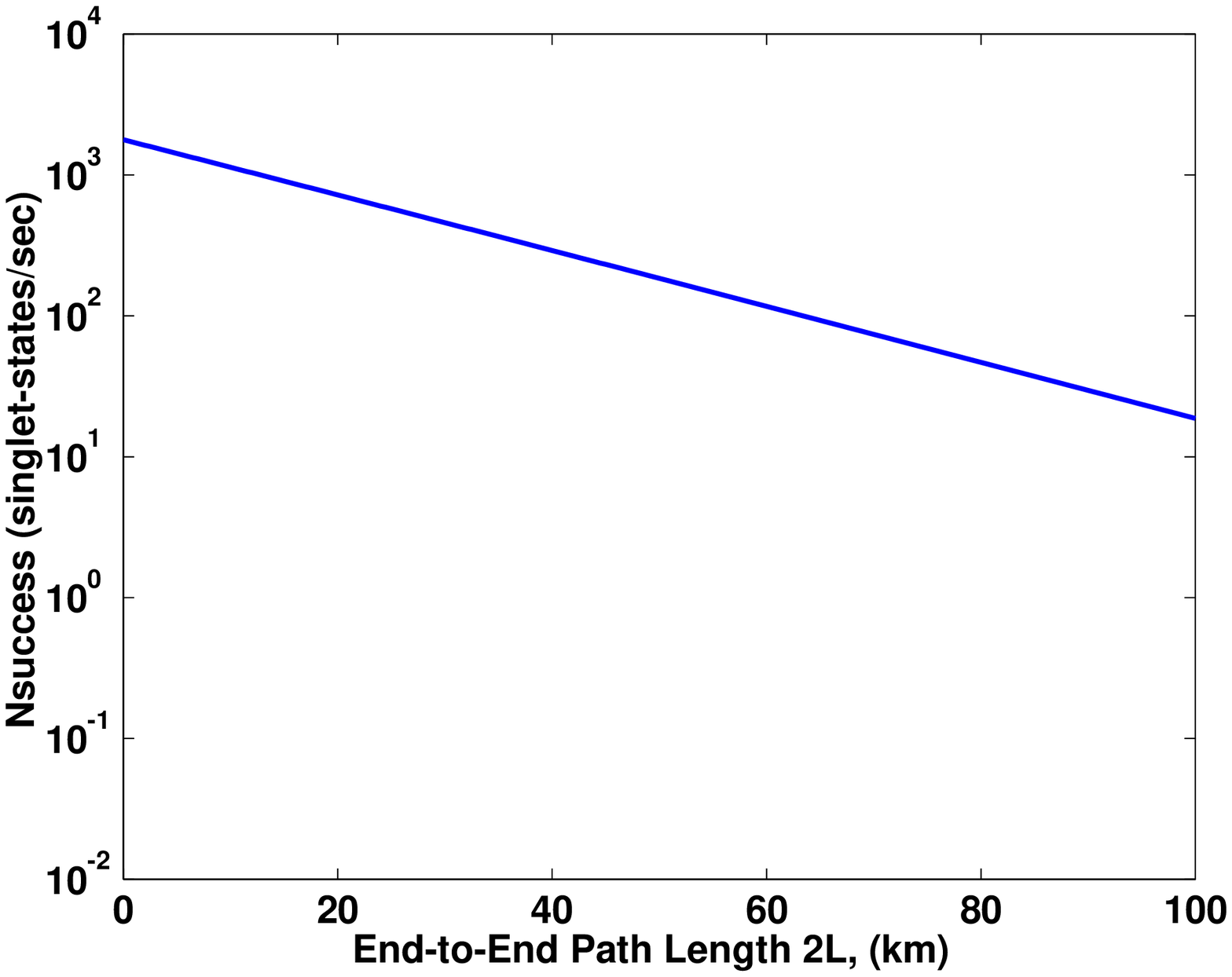}
\hspace*{.25in}\includegraphics[width=2.75in]{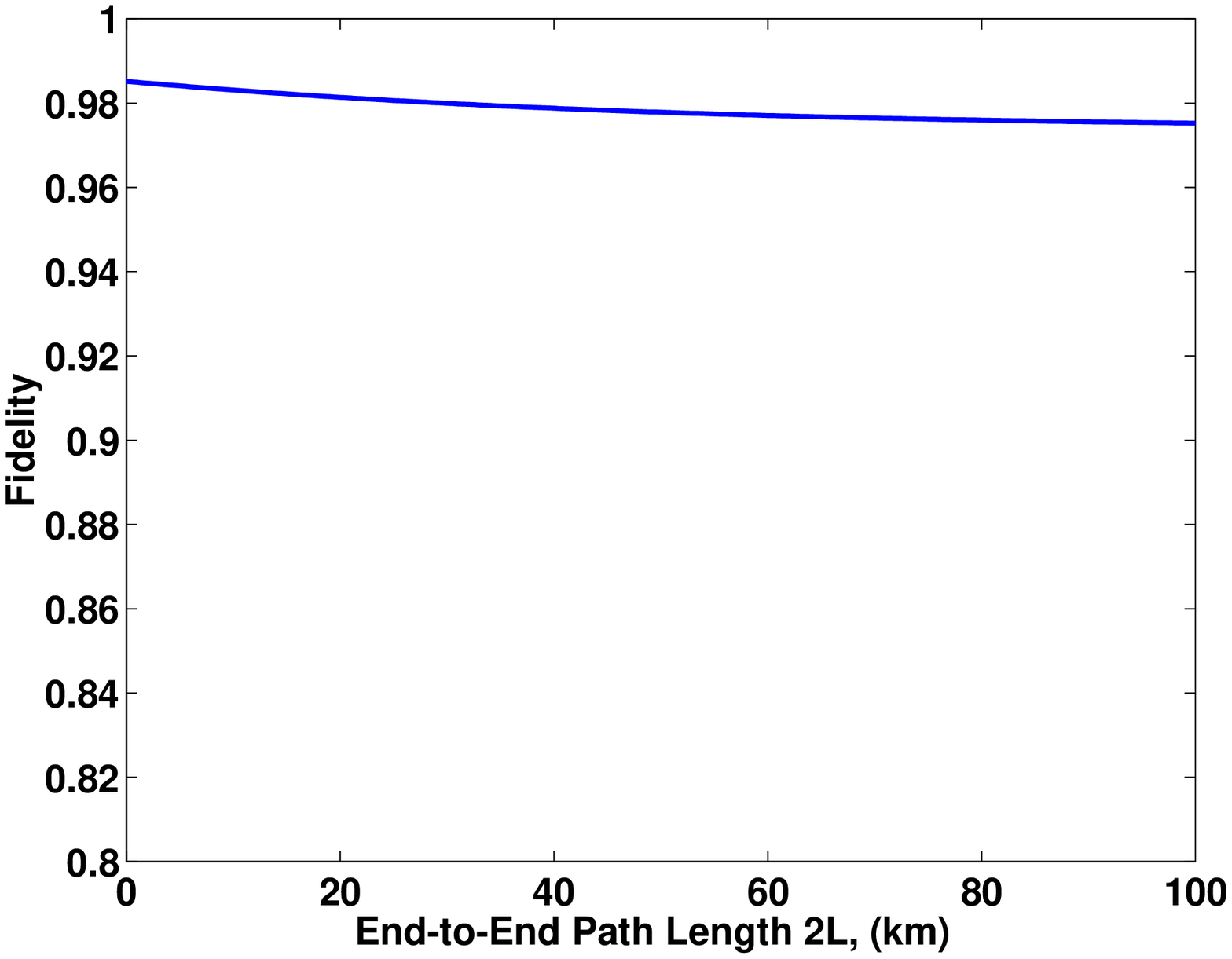}
\caption{\label{tel_perf} Figures of merit for the MIT/NU
teleportation architecture.  Left panel:  singlet-state throughput
versus end-to-end path length.  Right panel:  average teleportation
fidelity versus end-to-end path length.  Both panels assume the
following operating conditions:  dual-OPA source 
[Fig.~\ref{sourceandmemory}(b)] with each
OPA operated at 1\% of its oscillation threshold ($G^2 = 0.01$); 5\,dB of
excess loss in each
$P$-to-$M$ block path in Fig.~\ref{arch}; 0.2\,dB/km loss in each fiber;
$\Gamma_c/\Gamma = 0.5$ ratio of memory-cavity linewidth to source-cavity
linewidth; and $R = 500$\,kHz memory cycling rate.}
\end{figure}

The key figures of merit for our
teleportation system are its singlet-state throughput and its
average teleportation fidelity.  The throughput is the
average number of successful singlet-state loadings per second,
$N_{\text{success}} = R\Pr(\psi^{-})$, when the memory protocol is run at
rate $R$ and a lattice of Fig.~\ref{1_M_arch} 1:$K$-fanout quantum
memories is available for sequential loading at both the teleportation
transmitter and receiver.   In Fig.~\ref{tel_perf} we have plotted
these figures of merit versus the end-to-end path length between the
transmitter and the receiver under the following operating
conditions:  dual-OPA source [Fig.~\ref{sourceandmemory}(b)] with each
OPA operated at 1\% of its oscillation threshold ($G^2 = 0.01$); 5\,dB of
excess loss in each
$P$-to-$M$ block path in Fig.~\ref{arch}; 0.2\,dB/km loss in each fiber;
$\Gamma_c/\Gamma = 0.5$ ratio of memory-cavity linewidth to source-cavity
linewidth; and $R = 500$\,kHz memory cycling rate.  We see from this
figure that a throughput of nearly 200\,pairs/sec is achieved at an
end-to-end path length ($2L$) of 50\,km with an average teleportation
fidelity in excess of 97\%.
In the remainder of this section, we explore the use
of quantum error correction or entanglement purification to
increase the average teleportation fidelity shown in Fig.~\ref{tel_perf}.

\subsection{Quantum Error-Correcting Codes}

An $[[n,k,t]]$ quantum error correcting code is a mapping from $k$
logical qubits to $n$ physical qubits, with $k<n$, such that if $t$ or
fewer single-qubit errors occur on the physical qubits then we can
recover the original logical qubits perfectly.  A $[[5,1,1]]$ code
that saturates both the quantum Hamming bound and the quantum
Singleton bound was discovered by Laflamme et al.\@ \cite{laflamme}.  It
uses the following codewords,
\begin{align}
    |0_L\rangle &= |00000\rangle +|00110\rangle + |01001\rangle + 
|01111\rangle \notag\\
                & + |10101\rangle - |10011\rangle + |11100\rangle 
+|11010\rangle, \\
    |1_L\rangle &= -|00101\rangle -|00011\rangle +|01100\rangle 
-|01010\rangle \notag \\
                & - |10000\rangle + |10110\rangle +|11001\rangle + 
|11111\rangle.
\end{align}
Aung has shown \cite{jaung} that employing this five-qubit
code in the MIT/NU teleportation architecture results in a depolarizing
channel whose average teleportation fidelity satisfies,
\begin{equation}
    F = (2P_s' + 1)/3,
\end{equation}
where
\begin{equation}
    P_s' = (5 + 20P_s - 70P_s^2 + 40P_s^3 + 160P_s^4 - 128P_s^5)/27,
\end{equation}
gives the conditional probability of singlet loading \em
after\/\rm\ quantum error correction.

\begin{figure}
\includegraphics[width=2.75in]{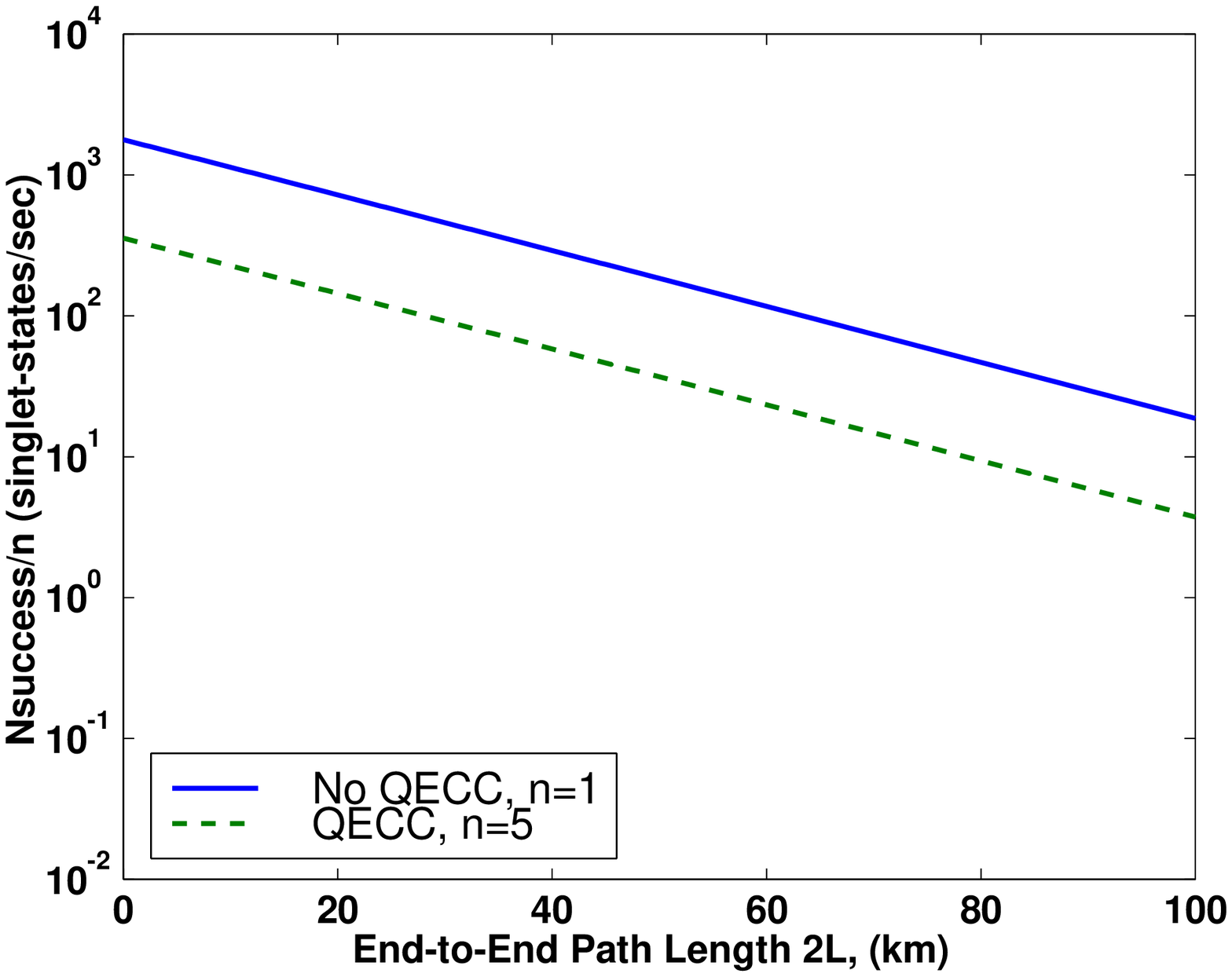}
\hspace*{.25in}
\includegraphics[width=2.75in]{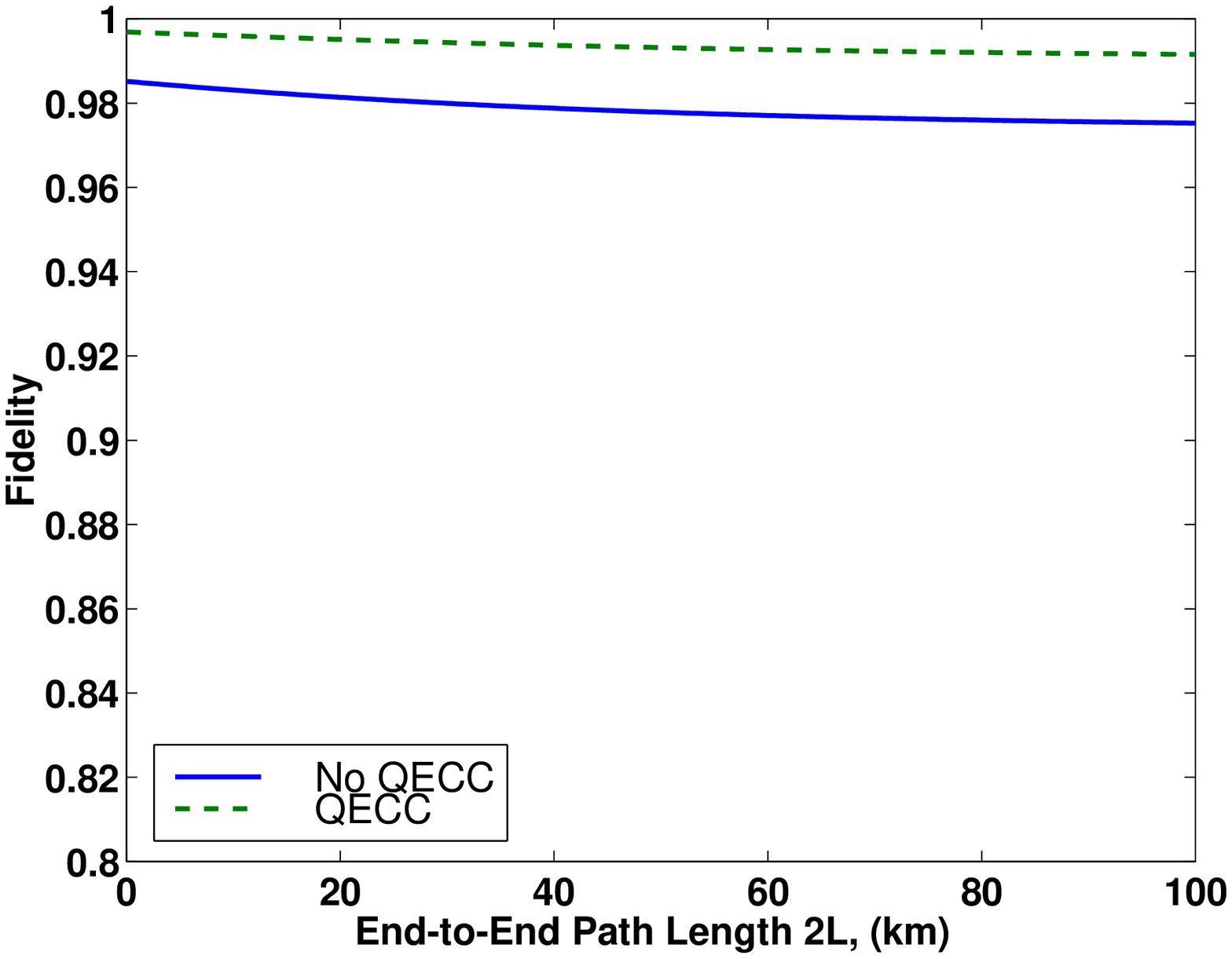}
\caption{\label{qecc_perf} Figures of merit for the MIT/NU
teleportation architecture with and without use of the five-qubit
quantum error-correcting code (QECC).  Left panel:  normalized throughput,
$N_{\text{success}}/n$, versus end-to-end path length.
Right panel:  average teleportation fidelity versus end-to-end path
length.  Both panels assume the same operating conditions as in
Fig.~\ref{tel_perf}.}
\end{figure}

Figure~\ref{qecc_perf} compares the performance of the teleportation
system when the five-qubit quantum error-correcting code (QECC) is
employed to the uncoded performance of this same system.  The operating
parameters that are assumed are the same as in Fig.~\ref{tel_perf}.
The left panel shows the normalized throughput, $N_{\text{success}}/n$,
versus end-to-end path length, where $n$ is the number of pairs used to
teleport a single message qubit.  Because the uncoded system employs
$n=1$ whereas the Laflamme et al. code requires $n=5$, this panel shows a
factor-of-five difference between the normalized throughputs of the
uncoded and coded systems.  The right panel of Fig.~\ref{qecc_perf} plots
average teleportation fidelity versus end-to-end path length.  This panel
shows the benefit of using the 5-qubit code:  $F
>0.99$ is now achieved out to a 100\,km end-to-end path length.

Further improvement in teleportation fidelity can be achieved by encoding
each five-qubit codeword using the five-qubit code.  The
result is a $[[n=25, k=1, t=3]]$ code, which indicates that we have lost
a factor of 25 in throughput to obtain the ability to correct all qubit
errors of order 3 or lower.  Although we could continue to
improve teleportation fidelity in this way,  it is clearly preferable to
consider quantum codes with larger block lengths that have better
minimum distance properties. Instead, we shall explore an
alternative---albeit closely related---technique known as entanglement
purification.

\subsection{Entanglement Purification Protocols}

An entanglement purification protocol (EPP) is a series of local
operations on $n$ entangled pairs designed to sacrifice $n-m$ of
these pairs in order to increase the fidelity of the $m$ remaining
pairs.  These protocols require classical communication between
locations sharing these pairs to coordinate the local operations,
i.e., communication between the transmitter and receiver of our
teleportation architecture.  In the limit of large block size $n$,
some protocols produce a finite number $m < n$ of near-perfect
singlets.  The yield of an entanglement purification protocol is
defined as $D = m/n$. Given a large number of entangled pairs and
protocol of yield $D$, we can use the resulting near-perfect
singlets for high-fidelity teleportation.

\begin{figure}
\includegraphics[width=2.75in]{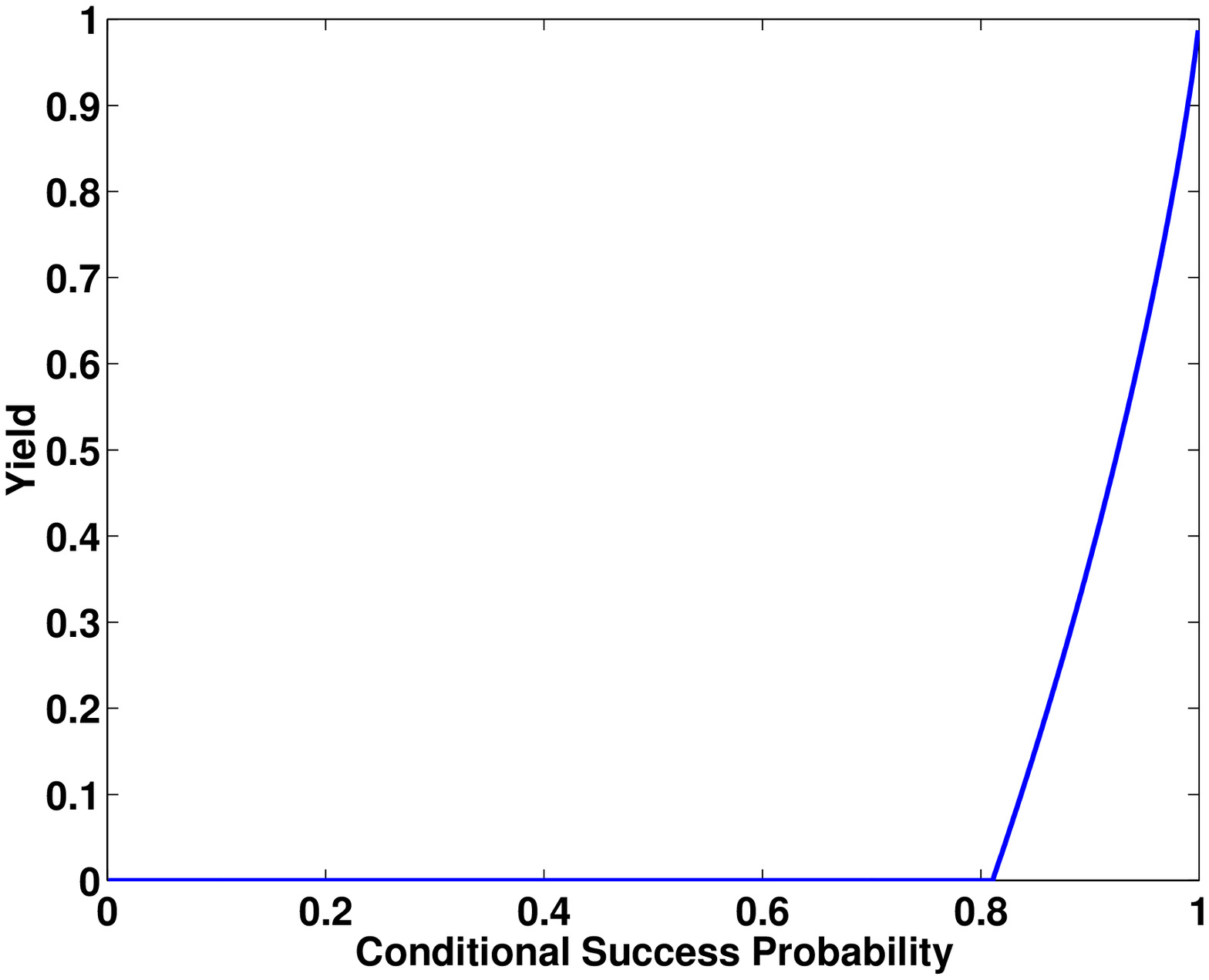}\\
\includegraphics[width=2.75in]{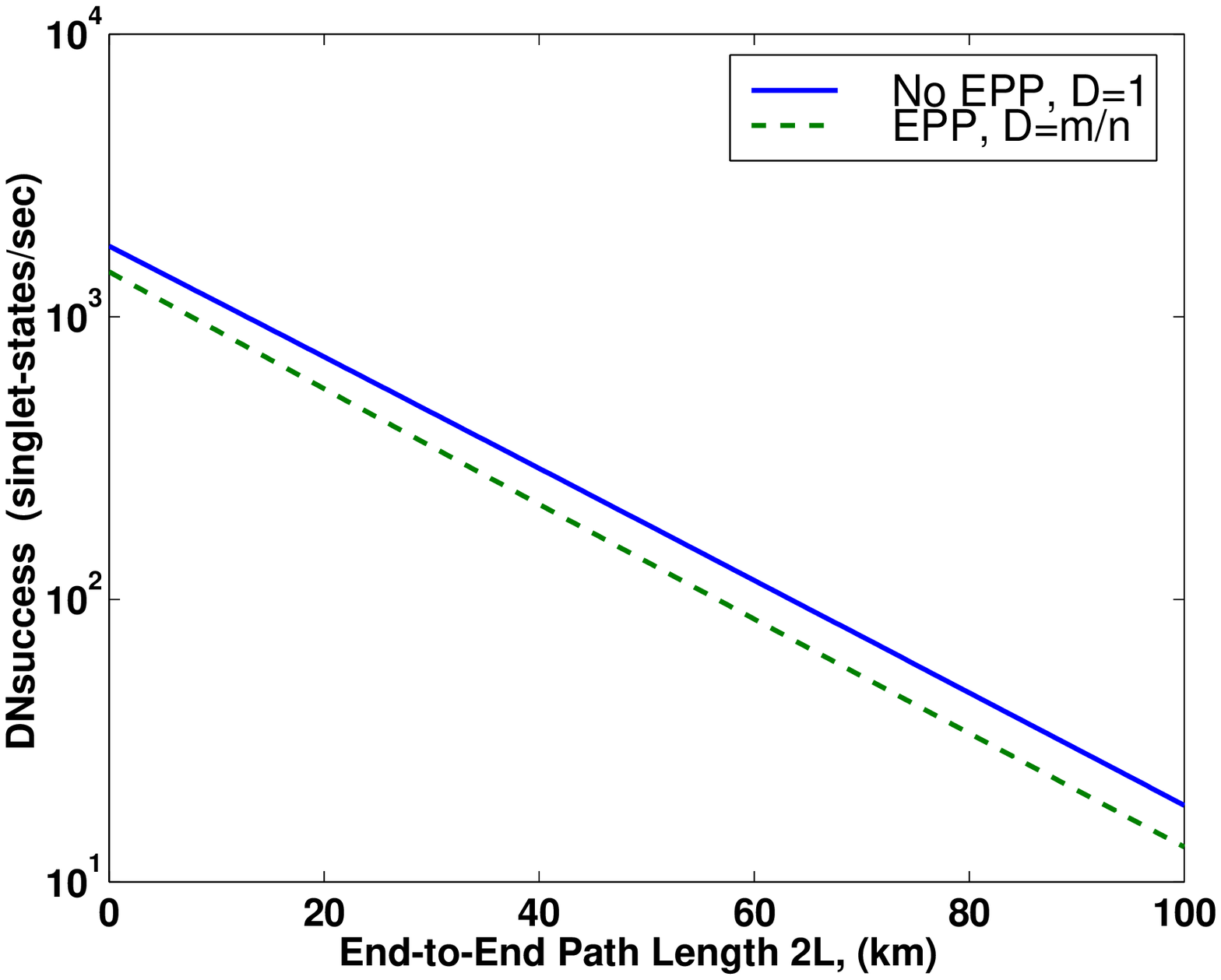}
\hspace*{.25in}
\includegraphics[width=2.75in]{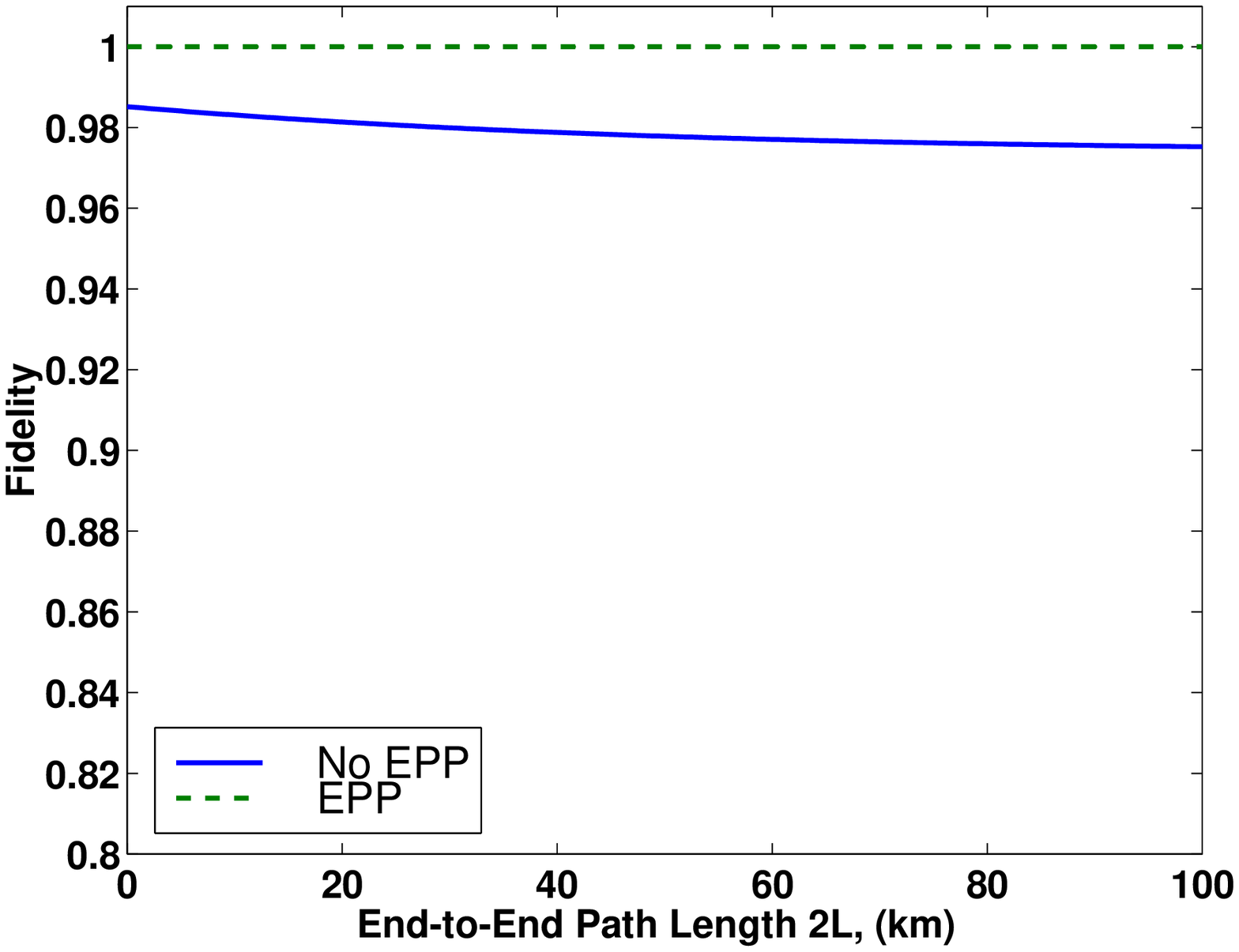}
\caption{\label{epp} Figures of merit for the MIT/NU
teleportation architecture with and without use of the one-way hashing
entanglement purification protocol (EPP).  Top panel:  EPP yield versus
conditional success probability, $P_s$.  Bottom left panel:  normalized
throughput,
$DN_{\text{success}}$, versus end-to-end path length.  Bottom right
panel:  average teleportation fidelity versus end-to-end path length.
All three panels assume the same operating conditions as in
Fig.~\ref{tel_perf}.}
\end{figure}

Bennett et al. proposed a one-way hashing EPP \cite{Bennett}.  For
$n\rightarrow \infty$  initial entangled pairs, with each pair in state
$\hat{\rho}$, they showed that the yield of their one-way hashing
protocol is $D = 1 - S(\hat{\rho})$ ideal singlets, where $S(\hat{\rho})
\equiv -\mbox{tr}[\hat{\rho}\log_2(\hat{\rho})]$ is the von Neumann
entropy of $\hat{\rho}$.  The MIT/NU architecture loads pairs that are
in the Werner state Eq.~\eqref{werner}, so if $n\gg 1$ pairs are stored
and the one-way hashing protocol is used we will get $Dn$ perfect
singlets where,
\begin{equation}
   D = 1 + P_s\log_2(P_s) + (1-P_s)\log_2[(1-P_s)/3].
\end{equation}
as plotted in the top panel of Fig.~\ref{epp}.  The yield is positive
only if the conditional success probability
satisfies
$P_s \geq 0.811$.  The uncoded MIT/NU architecture easily meets this
threshold out to 100\,km end-to-end path length.

The lower panels in Fig.~\ref{epp}
compare the performance of the MIT/NU teleportation architecture with and
without the use of the one-way hashing EPP.  The bottom left panel shows
normalized throughput, $DN_{\text{success}}$, versus end-to-end path
length, i.e., it plots the number of pairs/sec that will be available for
teleportation purposes
\em after\/\rm\ a yield-$D$ purification procedure has been employed,
where
$D=1$ when no purification protocol is employed.  Because the MIT/NU
architecture's initial fidelity is quite high, the one-way hashing
protocol has a high yield, so that throughput lost by virtue of employing
this EPP is quite modest.  However, because this EPP distills
perfect singlets---and because our performance analysis assumes perfect
Bell-state measurements at the transmitter and perfect qubit logic at the
receiver---the average teleportation fidelity with this protocol is
unity, as shown in the bottom right panel of Fig.~\ref{epp}.  There is a
substantial drawback, however, of the hashing protocol as compared to the
much simpler five-qubit QECC:  the EPP requires enormous amounts of
quantum memory at the transmitter and the receiver to realize the large
block sizes needed for validating use of the asymptotic yield expression
$D = 1 - S(\hat{\rho})$.

\section{GHZ-State Communication}

\begin{figure}
\includegraphics[width=2in]{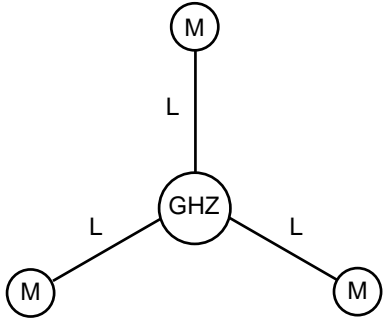}
\caption{\label{ghz} Schematic of long-distance GHZ communication
system. GHZ = source of polarization-entangled photons from
either the left or right panels of Fig.~\ref{ghzsources}; $L$ =
$L$\,km of standard telecommunications fiber; $M$ = trapped-atom quantum
memory.}
\end{figure}

There is considerable interest in Greenberger-Horne-Zeilinger
(GHZ) states \cite{GHZ}, because they can be used in a
non-statistical disproof of local hidden-variable theories of
physics and as resources for multiparty quantum communication
protocols such as quantum secret sharing \cite{hillary}.  As discussed in
\cite{Arch}, the MIT/NU teleportation architecture has an extension that
permits long-distance transmission and storage of three-party GHZ states,
\begin{equation}
    |\psi_{\text{GHZ}}\rangle = (|000\rangle +
    |111\rangle)/\sqrt{2},
\end{equation}
see Fig.~\ref{ghz}. In this section, we present the single-photon
loading event model for the GHZ-state quantum communication system, and
we use our model to quantify the performance achieved in quantum secret
sharing.

\subsection{GHZ-State Systems}

\begin{figure}
\includegraphics[width=2in]{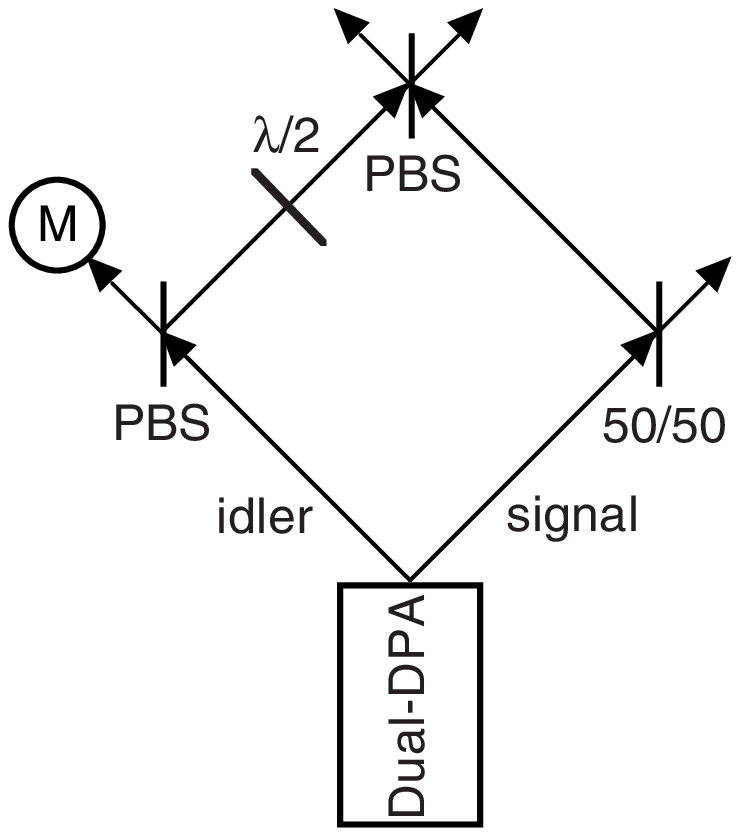}
\hspace*{.75in}\includegraphics[width=2in]{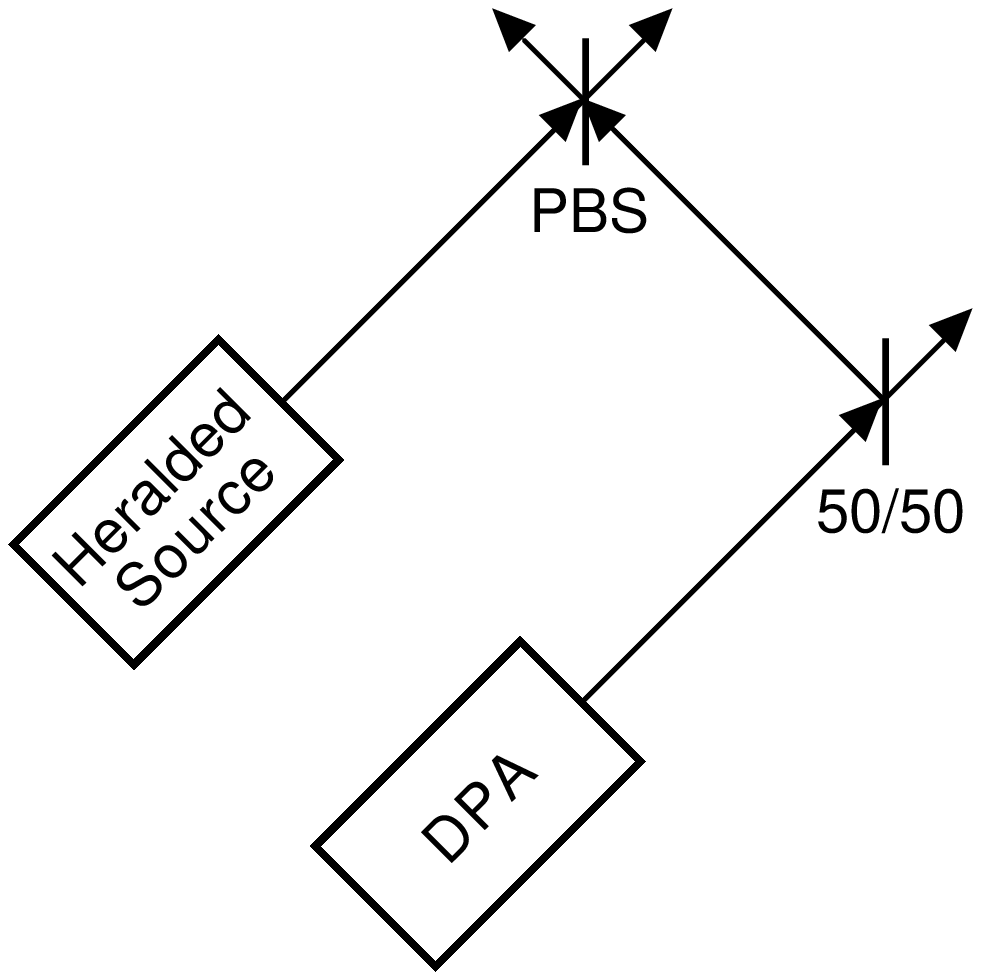}
\caption{\label{ghzsources} Source arrangements for the GHZ-state
communication architecture in Fig.~\ref{ghz}.  Left panel:  dual-DPA GHZ
system.  The quantum memory in this figure represents a memory
internal to the source block in Fig.~\ref{ghz}; its loading is used as a
trigger signal, see \cite{Arch} for details.  Right panel:  heralded
single-photon source plus DPA system. PBS = polarizing beam splitter,
$\lambda/2$ = half-wave plate.}
\end{figure}

The GHZ system in Fig.~\ref{ghz} is run under a clocked loading
protocol similar to the one described for singlet-state
transmission.  We employ quantum-state frequency conversion and
time-division multiplexing polarization restoration to
ensure that this GHZ-state quantum communication system is compatible with
transmission over standard telecommunications fiber.  We consider two
possible source arrangements for the GHZ block
in Fig.~\ref{ghz}.  The first is an ultrabright, narrowband variant of
the source used by Bouwmeester et al. in an initial experimental
demonstration of GHZ-state generation
\cite{bouwghz}.  That experiment was an annihilative table-top
measurement and had extremely low flux: 1~GHZ~state every 150\,sec.  Our
version of the Bouwmeester et al. source---shown in the left panel of
Fig.~\ref{ghzsources}---replaces their parametric downconverter with a
pair of doubly-resonant, type-II phase matched degenerate optical
parametric amplifiers (DPAs).  With this source, it was shown in
\cite{Arch} that the Fig.~\ref{ghz} arrangement permits a
throughput comparable to what
Bouwmeester et al. produced in the laboratory to be realized at a
source-to-memory radius of 10\,km. More important, though, is the fact
that the memories in the Fig.~\ref{ghz} architecture allow the GHZ state
to be stored for use in applications of three-party entanglement.

Recent work has shown that it may be possible to construct heralded
single-photon sources \cite{santori}.  With such a source, we can
design a GHZ system with a substantially higher throughput than
the configuration discussed above \cite{Arch}.  In the right
panel of Fig.~\ref{ghzsources}, the heralded source places a single
photon in the proper spatio-temporal mode for coupling to the trapped-atom
quantum memory during each loading cycle.  With the heralded-plus-DPA GHZ
source, throughput rises by three orders of magnitude over the
dual-DPA system, to about 15 GHZ states/sec at a 10\,km
source-to-memory radius.
\cite{Arch}.

\subsection{Single-Photon Event Models}

In this section, we present the single-photon loading event models for
the dual-DPA and heralded-plus-DPA GHZ-state quantum communication
systems.  We assume that each memory shown in Figs.~\ref{ghz} and
\ref{ghzsources} contains a 1:$K$-fanout array, as in
Fig.~\ref{1_M_arch}.  Consequently, we will only consider single-photon
loading events at each memory in Figs.~\ref{ghz} and \ref{ghzsources}.
Let $A$, $B$, and $C$ represent a clockwise labeling of the
memories in Fig.~\ref{ghz} starting from the lower left.  We define the
computational basis for these quantum memories to be,
\begin{align}
|0\rangle_A &= |01\rangle_{A_xA_y}\quad\mbox{and}\quad
|1\rangle_A =|10\rangle_{A_xA_y},\\
|0\rangle_B &= |01\rangle_{B_xB_y}\quad\mbox{and}\quad
|1\rangle_B = |10\rangle_{B_xB_y},\\
|0\rangle_C &= |10\rangle_{C_xC_y}\quad\mbox{and}\quad
|1\rangle_C =|01\rangle_{C_xC_y},
\end{align}
in terms of the number-ket representations for the
$x$- and $y$-polarized photons that loaded these memories. With this
computational basis, the GHZ state loaded by the Fig.~\ref{ghz} system is
$|\psi_{\text{GHZ}}\rangle_{ABC} =
(|000\rangle_{ABC}+|111\rangle_{ABC})/\sqrt{2}.$

It is not hard, using the basis,
\begin{equation}
\label{basis}
   \left\{ \frac{|000\rangle_{ABC} \pm |111\rangle_{ABC}}{\sqrt{2}},
      |001\rangle_{ABC}, |110\rangle_{ABC},
      |010\rangle_{ABC}, |101\rangle_{ABC},
      |011\rangle_{ABC}, |100\rangle_{ABC} \right\},
\end{equation}
to compute the matrix elements of the joint conditional density operator
for memories $A$, $B$, and $C$, given that an erasure has not occurred.
The derivation follows the approach that was used for the teleportation
architecture in Section~\ref{singlephoton}, so we shall omit
the details and just state the final results.  The conditional density
matrices for both the dual-DPA GHZ system and the heralded-plus-DPA GHZ
system turn out to be diagonal in the Eq.~\eqref{basis} basis.
For the dual-DPA source we find that,
\begin{equation}
   \label{ghzerrormodel}
   \hat{\rho}_{ABC} = \mbox{diag}\!
   \begin{pmatrix}
  P_{G_d} & 0 & P_{e1_d} & P_{e1_d} & P_{e1_d} & P_{e1_d} & P_{e2_d} &
P_{e2_d}
   \end{pmatrix},
\end{equation}
where
\begin{align}
P_{G_d} &=
\frac{(A^2+\tilde{n}^2)^2}{7A^4 + 12A^2\tilde{n}^2
+\tilde{n}^4},\\[.12in]
P_{e1_d} &=
\frac{A^2(A^2+2\tilde{n}^2)}{7A^4 + 12A^2\tilde{n}^2
+\tilde{n}^4},\\[.12in]
P_{e2_d} &=
\frac{A^2(A^2+\tilde{n}^2)}{7A^4 + 12A^2\tilde{n}^2
+\tilde{n}^4},
\end{align}
with $A\equiv \bar{n}(1+\bar{n})-\tilde{n}^2$.
For the heralded-plus-DPA source we get,
\begin{equation}
   \label{errormodelherald}
   \hat{\rho}_{ABC} =
   \mbox{diag}\!\begin{pmatrix}
         P_{G_h} & 0 & P_{e1_h} & P_{e1_h} & P_{e2_h} & 0 & P_{e2_h} & 0
   \end{pmatrix},
\end{equation}
where
\begin{align}
P_{G_h} &=
\frac{\eta(A^2 + \tilde{n}^2)[(1+\bar{n})^2-\tilde{n}^2]}
{\eta(3A^2+\tilde{n}^2)[(1+\bar{n})^2-\tilde{n}^2]
+2(1-\eta)A(A^2+2\tilde{n}^2)},\\[.12in]
P_{e1_h} &=
\frac{\eta A^2[(1+\bar{n})^2-\tilde{n}^2]}
{\eta(3A^2+\tilde{n}^2)[(1+\bar{n})^2-\tilde{n}^2]
+2(1-\eta)A(A^2+2\tilde{n}^2)},\\[.12in]
P_{e2_h} &=
\frac{(1-\eta)A(A^2+2\tilde{n}^2)}
{\eta(3A^2+\tilde{n}^2)[(1+\bar{n})^2-\tilde{n}^2]
+2(1-\eta)A(A^2+2\tilde{n}^2)}.
\end{align}
In calculating
these matrix elements we have used the same transmission loss factor,
$\eta =
\eta_L\gamma\gamma_c/\Gamma\Gamma_c$, for each source-to-memory path in
Figs.~\ref{ghz} and \ref{ghzsources}.

\subsection{Performance Analysis}

Secret sharing refers to cryptographic protocols which allow Alice
to share secret information with Bob and Charlie in such a way
that individually they have no means for learning Alice's secret,
but by working together can they gain access to Alice's secret
information. One classical implementation of secret sharing
requires Alice to send a random bit string $r$ to Bob and the
modulo-2 sum, $r\oplus m$, of the random bit string $r$ and her
message $m$ to Charlie.  If Bob and Charlie act together, they can
recover Alice's message $m$ simply by adding their bit strings
together.

We consider the performance of our GHZ systems in the quantum
secret sharing (QSS) protocol proposed in Ref.~\cite{hillary}. In
this protocol, Alice, Bob, and Charlie share a GHZ state $|\psi_{\rm
GHZ}\rangle_{ABC}  = (|000\rangle_{ABC} +
    |111\rangle_{ABC})/\sqrt{2}$, and Alice's secret is the qubit
$|\psi\rangle_S =
\alpha|0\rangle_S + \beta|1\rangle_S$, which she wishes to send to
Bob and Charlie in such a way that they must cooperate to obtain
this quantum information.  The joint state of Alice, Bob, and
Charlie---including Alice's portion of the GHZ state \em and\/\rm\ her
quantum secret---at the start of the QSS protocol is
$|\psi\rangle_S|\psi_{\rm GHZ}\rangle_{ABC}$.

Alice initiates the QSS protocol by making the Bell-state measurements,
$\{ |\psi^\pm\rangle_{SA}, |\phi^\pm\rangle_{SA}\}$, on her secret and her
portion of the GHZ state.  Alice then labels $(m,n)$, the two classical
bits she derives from these measurements, using the following scheme:
$\psi^+=(0,1), \psi^-=(1,1), \phi^+=(0,0),\phi^-=(1,0)$.  She sends
$m$ to Bob and
$m\oplus n$ to Charlie, using secure classical channels so that Bob
cannot intercept $m\oplus n$ and Charlie cannot obtain $m$.  It
follows that neither Bob nor Charlie has any information about Alice's
secret---even after receiving the classical information from
Alice---because their marginal density operators can be shown to be
$\hat{\rho}_B = \hat{I}_B/2$ and $
\hat{\rho}_C = \hat{I}_C/2$, respectively, where $\hat{I}$ is the
identity operator, at this point in the protocol.

For Bob and Charlie to learn Alice's secret qubit $|\psi\rangle_S$, they
must cooperate.  Because the no-cloning theorem precludes making two
copies of this state, either Bob or Charlie---but \em not\/\rm\ both of
them---will possess a replica of $|\psi\rangle_S$ at the end of the QSS
protocol.  Let us arbitrarily assume that Bob and Charlie have agreed to
let Charlie be the recipient of this replica.  Having made that
agreement, Bob measures his portion of the GHZ state in the $x$ basis,
$\{|\pm x\rangle_B \equiv (|0\rangle_B\pm|1\rangle_B)/\sqrt{2}\}$, and
he sends Charlie the result of this measurement along with Alice's
$m$ bit.  Together with Alice's $m\oplus n$---which he
received earlier---Charlie now has all the information he needs to turn
his portion of the GHZ state into a replica of Alice's secret via a
local unitary operation.

Let $F$ be the
average fidelity of the preceding QSS protocol when Alice's secret,
$|\psi\rangle_S$,  is selected from a uniform distribution over the Bloch
sphere.  From our single-photon event models, it can be shown that the
average QSS fidelity for the dual-DPA GHZ system is,
\begin{equation}
    F = P_{G_d} + 2P_{e1_d} + 2P_{e2_d}/3,
\end{equation}
and the average QSS fidelity for the heralded-plus-DPA GHZ system is,
\begin{equation}
    F = P_{G_h} + 2P_{e1_h}/3 + P_{e2_h}.
\end{equation}

\begin{figure}
\includegraphics[width=3.75in]{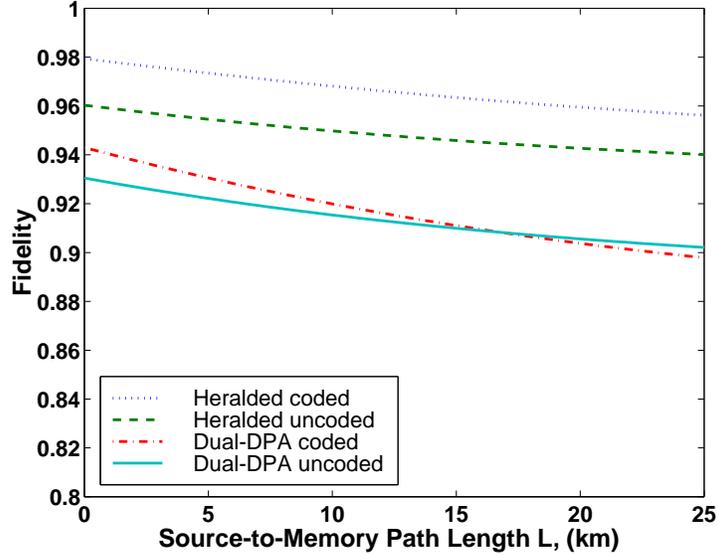}
\caption{\label{qssfid} Average fidelity in the QSS protocol. We
compare the performance of the dual-DPA and heralded GHZ systems
with and without coding.  These plots assume each DPA operates at 1\% of
its oscillation threshold, 5\,dB excess loss in
each source-to-memory path, 0.2 dB/km loss in each fiber, and
$\Gamma_c/\Gamma = 0.5$ ratio of memory-cavity linewidth to
source-cavity linewidth.}
\end{figure}

Quantum error correction can be used to improve the performance of
the QSS protocol.  For the five-qubit error-correcting code
from \cite{laflamme} we used simulations to calculate the QSS performance
of our GHZ systems.  Figure~\ref{qssfid} shows the average QSS fidelity
for the dual-DPA and heralded-plus-DPA GHZ systems with and without
coding.  We see that the heralded-plus-DPA GHZ system has significantly
better performance than the dual-DPA system in the QSS protocol in both
uncoded and coded operation.  Coding improves the performance of the
heralded-plus-DPA system for all path lengths shown in this figure, but
beyond about 16\,km source-to-memory path length coding reduces the
fidelity of the dual-DPA system.  The dual-DPA curves with and without
error correction cross because the five-qubit code degrades performance
when the incidence of multi-qubit errors is too high; the same thing
occurs for the heralded-plus-DPA system, but at a much longer path
length.

\section{Conclusions}

We have reviewed the MIT/NU quantum communication architecture for
long-distance, high-fidelity qubit teleportation and developed the
single-photon loading event model for this system.  Using this model, we
have quantified the fidelity improvements offered by a simple quantum
error-correcting code, and by a powerful entanglement purification
protocol.  The MIT/NU architecture has an extension that enables
long-distance transmission and storage of Greenberger-Horne-Zeilinger
states; we have derived the single-photon loading event model for this
system too and examined its performance with and without the use of a
simple quantum error-correcting code.

\begin{acknowledgments}
This research was supported by the DoD Multidisciplinary University
Research Initiative (MURI) program administered by the Army
Research Office under Grant DAAD 19-00-1-0177, by the Quantum
Information Science and Technology program under Army Research
Office Grant DAAD-19-01-0647, and by the National Reconnaissance
Office under Contract NRO-000-C-0158.
\end{acknowledgments}

\bibliography{Shapiro_Aung_Yen_final}

\end{document}